\documentclass[aip,apl,preprint]{revtex4-1}
\usepackage{graphicx}% Include figure files
\usepackage{dcolumn}% Align table columns on decimal point
\usepackage{bm}% bold math

\begin{document}

%\preprint{APS/123-QED}

\title{Independent individual addressing of multiple neutral atom qubits with a MEMS beam steering system}% Force line breaks with \\

\author{C. Knoernschild}
\affiliation{Fitzpatrick Institute for Photonics, Electrical and Computer Engineering Department, Duke University, Durham, North Carolina 27708, USA}

\author{X. L. Zhang}
\author{L. Isenhower}
\author{A. T. Gill}
\affiliation{Department of Physics, 1150 University Avenue, University of Wisconsin, Madison, Wisconsin 53706, USA}

\author{F. P. Lu}
\affiliation{Applied Quantum Technologies, Durham, NC, 27707}

\author{M. Saffman}
\affiliation{Department of Physics, 1150 University Avenue, University of Wisconsin, Madison, Wisconsin 53706, USA}

\author{J. Kim}%
\email{jungsang@duke.edu}
\affiliation{Fitzpatrick Institute for Photonics, Electrical and Computer Engineering Department, Duke University, Durham, North Carolina 27708, USA}%
\affiliation{Applied Quantum Technologies, Durham, NC, 27707}

\date{\today}% It is always \today, today,

\begin{abstract}

We demonstrate a scalable approach to addressing multiple atomic qubits for use in quantum information processing. Individually trapped $^{87}$Rb atoms in a linear array are selectively manipulated with a single laser guided by a MEMS beam steering system. Single qubit oscillations are shown on multiple sites at frequencies of $\simeq3.5$ MHz with negligible crosstalk to neighboring sites. Switching times between the central atom and its closest neighbor were measured to be $6$-$7$ $\mu$s while moving between the central atom and an atom two trap sites away took $10$-$14$ $\mu$s. 

\end{abstract}

\maketitle

Scalable quantum information processing (QIP) has become a serious topic of consideration over the past decade. Among the various physical systems considered for scalable QIP implementation, individually trapped neutral atoms or ions serving as quantum bits (qubits) have emerged as leading candidates. Both of these approaches use precisely tuned lasers or microwave radiation to perform single and multi-qubit gate operations\cite{LeibfriedN2003,Schmidt-KalerN2003,IsenhowerPRL2010,WilkPRL2010} and have shown the ability to trap and image multiple atomic qubits\cite{SeidelinPRL2006,NagerlAPB1998,PortoRS2003,NelsonNP2007,LengwenusAPB2007}. In order to extend these quantum gate operations over a larger array of qubits, a scalable way of distributing laser or microwave resources is necessary. Previously, this has been accomplished using magnetic field gradients\cite{SchraderPRL2004,WangAPL2009} or acousto/electro-optic deflectors\cite{YavuzPRL2006,Schmidt-KalerAPB2003}. In this letter we demonstrate a more scalable laser multiplexer utilizing microelectromechanical systems (MEMS) technology, which can easily be extended to multiple laser beams across a range of wavelengths providing the capability to address multiple trap sites simultaneously.\cite{KnoernschildOL2008,KnoernschildOE2009} We report a significant step toward scalable QIP in atomic lattices by incorporating a two dimensional (2D) MEMS laser steering system into a multi-site neutral atom trap experiment. Raman-induced single qubit rotations between hyperfine ground states are shown on individual $^{87}$Rb atoms in a linear array of $5$ trap sites with negligible impact on neighboring atoms along with a quantitative characterization of the Rabi frequency and switching times achieved in our setup.

%\begin{figure}
%\includegraphics{MEMS}% Here is how to import EPS art
%\caption{\label{fig:mems} (a) Schematic of the folded imaging system used to combine orthogonal tilt directions of MEMS mirrors. Each MEMS mirror reflects the beam twice to double the scanning range of the system. (b) Scanning electron micrograph of a typical MEMS mirror used in the system.}
%\end{figure}

\begin{figure}
\includegraphics[width=8.5cm]{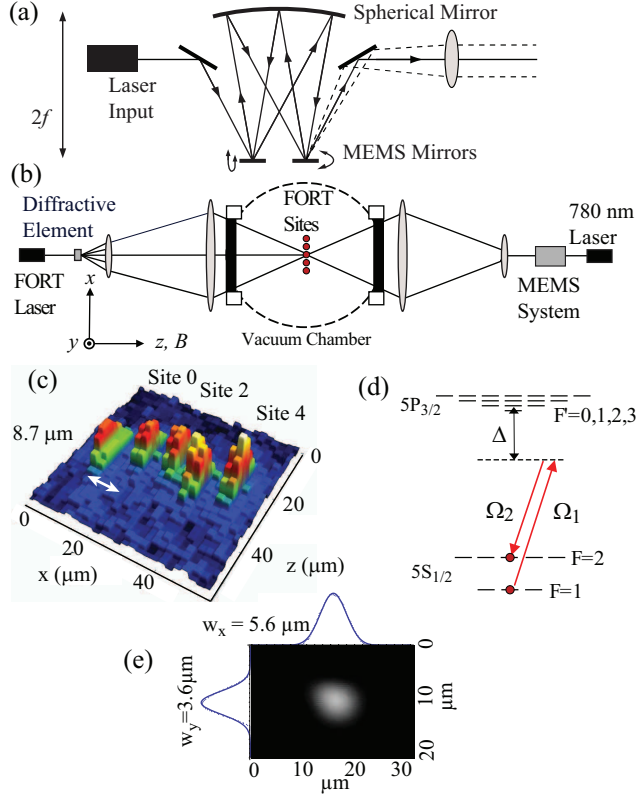}% Here is how to import EPS art
\caption{\label{fig:setup} (a) MEMS steering system schematic. (b) Schematic of the experimental setup showing the FORT beams and the $780$ nm ground state Rabi laser. (c) Averaged fluorescence image of the $5$ trap sites over many trap loading periods. Each site is separated by $8.7$ $\mu$m. (d) Energy level structure showing the two photon Raman transition used for single qubit rotations. (e) Image and profile of the focused ground state Rabi laser at the atom.}
\end{figure}

The design of the 2D MEMS steering system, shown in Fig. \ref{fig:setup}(a), is guided by fast steering speed requirements. Beam steering is accomplished with two 1D tilting micromirrors designed to achieve a high mechanical resonant frequency while maintaining near critical damping to minimize settling time when switching between sites.~\cite{ChangsoonSTiQEIJo2007} The micromirrors are fabricated on the same substrate, using a commercial foundry process (PolyMUMPS, offered by MEMSCAP, Inc.~\cite{Memscap}), and designed to tilt in orthogonal directions. Each micromirror consists of a gold coated polysilicon plate with a radius of $100$ $\mu$m that rotates about two torsional beams. The micromirror is electrostatically tilted to an angle $\theta$ by way of a potential difference between the grounded mirror plate and an applied voltage $V$ on the underlying actuation electrodes. A simple $2f$-$2f$ folded imaging system is used to optically combine the perpendicular steering axes. An angle-multiplication scheme is implemented by allowing the beam to reflect off each micromirror twice, which reduces the maximum angular tilt requirement to enable mirrors with higher resonant frequencies.\cite{KnoernschildOL2008}

Individual $^{87}$Rb atoms are spatially confined in far-off-resonant traps (FORTs) in a similar experimental setup to the one described in Ref. \onlinecite{UrbanNP2009}. The FORT sites are created by a tightly focused $1064$ nm laser beam split with a diffractive element into $5$ spots each with a beam waist of $3.4$ $\mu$m [Fig. \ref{fig:setup}(b)]. This produces a linear array of $5$ trap sites spatially separated by $8.7$ $\mu$m with a peak potential depth of $\sim4.5$ mK. The atoms are loaded into the FORT sites from an atom cloud confined by a magneto-optical trap (MOT). Loading of the sites is checked by collecting resonant fluorescence from the atoms on an electron-multiplying charge-coupled device camera and integrating counts over regions of interest aligned with each of the 5 FORT sites. The atoms are then Doppler cooled with the MOT beams to temperatures of $200$-$250$ $\mu$K. Figure \ref{fig:setup}(c) shows an averaged fluorescence image of many single atom loading periods with sufficient spacing to resolve individual atoms in each lattice site.

The qubit is defined by two magnetic field insensitive hyperfine ground states $\left|0\right\rangle\equiv\left|F=1,m_F=0\right\rangle$ and $\left|1\right\rangle \equiv \left|F=2,m_F=0\right\rangle$. Single qubit rotations between these two states are induced by a two photon Raman transition as shown in Fig. \ref{fig:setup}(d). This process uses a single ground state Rabi laser with two frequency components separated by the hyperfine splitting $6.8$ GHz generated by current modulation of a $780$ nm diode laser. The laser is detuned from the $5P_{3/2}$ excited state by $\Delta = -2\pi\times100$ GHz and is circularly polarized with respect to the magnetic field direction which is aligned along the $z$ axis. It propagates through the MEMS steering system, which is aligned to address the central atom (site $2$) when unactuated, and focuses onto the targeted atom. The beam at the atom is slightly elliptical with beam waists of $w_x = 5.6$ $\mu$m and $w_y = 3.6$ $\mu$m [Fig.\ref{fig:setup}(e)]. 

In our first experiment, initial states are prepared by applying a magnetic field along the z axis to split the $m_F$ energy levels and optically pumping the atoms into the $\left|1\right\rangle$ state. The Rabi laser is directed to a target atom for a duration of $T$. The resulting state is then measured by selectively ejecting the atoms in the $\left|1\right\rangle$ state from the trap and detecting fluorescence from the undisturbed atoms in the $\left|0\right\rangle$ state as described in Ref. \onlinecite{IsenhowerPRL2010}.

\begin{figure}
\includegraphics[width=8.5cm]{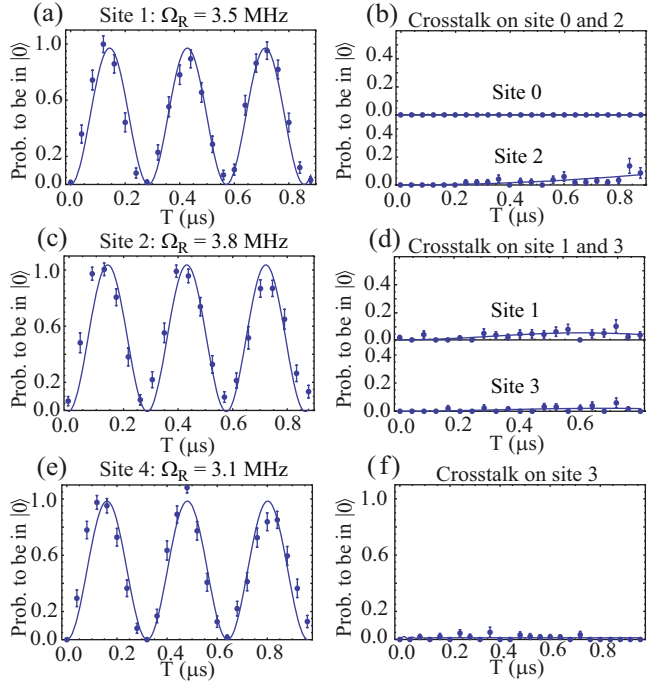}% Here is how to import EPS art
\caption{\label{fig:rabi} Rabi flopping on target atom and neighboring atom(s) when the laser addresses sites (a)-(b) $1$, (c)-(d) $2$, and (e)-(f) $4$.}
\end{figure}

The frequency of the ground state Rabi oscillations is given by $\Omega_R = \Omega_1 \Omega_{2}^{*} / 2\Delta$ where the Rabi frequencies of the two laser side bands are $\Omega_1$ and $\Omega_2$, and the detuning from the excited state $\Delta$ is much larger than the excited state decay rate. By measuring the probability of detecting the atom in the $\left|0\right\rangle$ state as a function of the Rabi laser duration $T$, we see coherent Rabi oscillations between the two qubit states at several of the $5$ trap sites [Fig. \ref{fig:rabi}]. When the MEMS system targets site $1$, the atom oscillates between the qubit states at a frequency of $3.5$ MHz and a contrast of $0.97\pm0.04$ [Fig. \ref{fig:rabi}(a)] while the crosstalk to neighboring atoms at sites $0$ and $2$ shown in (b) is negligible, within the experimental noise limit of the system. Figure \ref{fig:rabi}(c) [\ref{fig:rabi}(e)] shows the Rabi oscillation of the targeted site with a frequency of $3.8$ MHz ($3.1$ MHz) and constrast of $1.04\pm0.05$ ($0.98\pm0.04$) when the MEMS system addresses site $2$ (site $4$) while Rabi flopping at neighboring sites are shown in (d) [(f)]. The difference in Rabi frequencies indicates a variation in the laser intensity the atom experiences at each of the trap sites. The intensity difference among sites arises from the aberrations in the MEMS folded imaging system and small actuation-induced curvature of the micromirrors, but its effects can be calibrated out by adjusting the pulse duration at each site. This experiment demonstrates individual qubit addressability with negligible crosstalk with a MEMS based beam steering system.

\begin{figure}
\includegraphics[width=8.5cm]{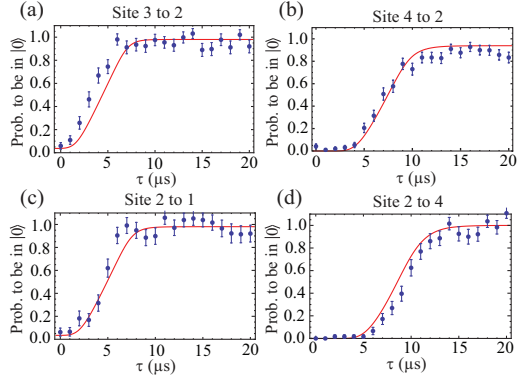}% Here is how to import EPS art
\caption{\label{fig:switch} Experimental data (points) and model (solid line) showing the probability the atom has switched from the $\left|1\right\rangle$ to the $\left|0\right\rangle$ state after a delay $\tau$ between the MEMS system trigger and a $\pi$-pulse for transitions from (a) $3\rightarrow2$, (b) $4\rightarrow2$, (c) $2\rightarrow1$, and (d) $2\rightarrow4$.}
\end{figure}

In addition to individual addressability, we examine the time it takes to switch the beam between atoms using the following procedure. First, the atom at the target site is prepared in the initial $\left|1\right\rangle$ state. Then, the MEMS system is triggered to adjust the beam path from an initial site to the target site at time $t=0$. After a delay of $\tau$, a timed pulse from the Rabi laser is used to flip the illuminated atom to the $\left|0\right\rangle$ state ($\pi$-pulse). The resulting atomic state is determined by state-selective measurement. Since the $\pi$-pulse duration ($\sim140$ ns) is sufficiently less than the switching time of the MEMS system ($\sim10$ $\mu$s), this measurement should provide an accurate indication of the fraction of the $\pi$-pulse acting on the atom for various delays $\tau$, leading to the determination of the switching time. Figure \ref{fig:switch} shows the probability of finding the target atom in the $\left|0\right\rangle$ state as a function of $\tau$ for several configurations. When the $\pi$-pulse is launched with short delay, the state of the atom does not make a complete transition to the $\left|0\right\rangle$ state as the beam has not arrived at the target location. When it is launched after sufficient delay, the full $\pi$-pulse impinges on the atom leading to a complete state change. The solid lines are theoretical models generated from independent transient measurements of the mirror tilt and the atomic rate equations that describe the transition probability. The model also takes into account a uniform timing jitter distribution in $\tau$ that arises from hardware limitations in the experiment ($\sim2$ $\mu$s). Switching times of $6$-$7$ $\mu$s are measured when the MEMS system moves the laser between site $2$ (center) and its neighboring sites while they increase to $10$-$14$ $\mu$s when addressing involves the outer sites. The switching is faster when the mirror is returning to site $2$ as shown in Fig \ref{fig:switch} by comparing (a) and (b) with (c) and (d), respectively. This is due to well-known electrostatic softening effect in MEMS actuation\cite{ChangsoonSTiQEIJo2007}, which results in a slower response as the MEMS system (1) addresses sites further from the center and (2) switches from the middle site to one of the outer sites compared to when it returns back to the central site. We note the actual transition time of the atomic state is less than $5$ $\mu$s, in part due to timing jitter in the present experimental setup. The net delay can be compensated by pre-triggering if multiple beams are available in the MEMS system.

In summary, this letter has demonstrated individual addressability of a linear array of $5$ trapped $^{87}$Rb atoms with negligible cross talk to neighboring atoms using a scalable MEMS technology. Rabi frequencies at these sites of $3.1-3.8$ MHz show the capability of fast single qubit gates. MEMS steering technology promises scalability to address more trap sites using more simultaneous lasers across a wide range of wavelengths at adequate switching speeds. While we have focused on its use in neutral atom QIP, it is applicable to other qubit systems in a periodic lattice including atomic ions, diamond nitrogen-vacancy color centers, and quantum dots.

This work is supported by ARO under contract W911NF-08-C-0032, ARO/IARPA under contract W911NF-05-1-0492, and NSF grant PHY-0653408.

\clearpage

%\bibliography{mybib}% Produces the bibliography via BibTeX.

%merlin.mbs 2010-03-15 4.21a (PWD, AO, DPC)

\end{document}